\documentstyle[11pt,aaspp]{article}

\def\etal{et~al.\ }
\def\eg{{\it e.g.\ }}
\def\ie{{\it i.e.\ }}
\def\cf{{\it cf.\ }}

\def\minmag{\lower.5ex\hbox{$\; \buildrel < \over > \;$}}
\def\magmin{\lower.5ex\hbox{$\; \buildrel > \over < \;$}}
\def\gtwid{\mathrel{\raise.3ex\hbox{$>$\kern-.75em\lower1ex\hbox{$\sim$}}}}
\def\ltwid{\mathrel{\raise.3ex\hbox{$<$\kern-.75em\lower1ex\hbox{$\sim$}}}}
\def\ref{\par\noindent\hangindent.5in\hangafter=1}
\def\kms{\hbox{km s$^{-1}$}}
\def\percmsq{\hbox{cm$^{-2}$}}
\def\ergs{\hbox{erg s$^{-1}$}}
\newcommand{\beq}{\begin{equation}}
\newcommand{\eneq}{\end{equation}}
\newcommand{\beqar}{\begin{eqnarray}}
\newcommand{\eneqar}{\end{eqnarray}}
\newcommand{\barn}{\begin{eqnarray*}}
\newcommand{\earn}{\end{eqnarray*}}

\lefthead {Krolik \& Kriss}
\righthead {Properties of Warm Gas}

\begin{document}

\title{Observable Properties of X-ray Heated Winds in AGN:
Warm Reflectors and Warm Absorbers}

\author{Julian H. Krolik}

\and

\author{Gerard A. Kriss}

\affil{Department of Physics and Astronomy, Johns Hopkins University,
	   Baltimore, MD 21218}

\begin{abstract}

    First discovered by spectropolarimetry, the warm reflecting gas near
active galactic nuclei may be observed in many ways.  When the nucleus
itself is obscured, and instrumental angular resolution is fine enough
to exclude radiation from the host galaxy, this gas can be seen in
the soft X-ray band by a combination of bremsstrahlung, intrinsic
line emission, and reflection of the nuclear continuum, both by electron
scattering and by resonance line scattering.  Strong blended line features
can be expected in the spectrum.
We show that the strong emission line features
seen in the keV region in the X-ray spectra of obscured AGN may be
due to this gas, partly due to intrinsic emission (as suggested by some
previous studies), and partly due to resonance scattering.
In the ultraviolet, intrinsic emission
is very weak, and strongly dominated by lines.  Reflection, principally
by electron scattering, but also with some contribution from resonance lines,
is the main signature in the UV.

  When our line of sight to the nucleus is not
obscured, the dominant effect is absorption.  In the soft X-ray band,
ionization edges of highly ionized species and resonance lines contribute
comparably to the opacity; in the ultraviolet, the gas
is almost transparent except for a small number of resonance lines.

    We identify the ``warm absorbers" seen in many AGN X-ray spectra
with this gas, but argue that most of the UV absorption lines seen must
be due to a small amount of more weakly ionized gas which is embedded in the
main body of the warm, reflecting gas.  Because the ionization
equilibration timescales of some ions may be as long as the variability
timescales in AGN, the ionic
abundances indicated by the transmission spectra may not be well-described by
ionization equilibrium.
\end{abstract}

\section{Introduction}

    The first strong evidence that significant quantities of warm ($T \sim
10^6$K) gas could be found near active galactic nuclei came from the
spectropolarimetry of Antonucci \& Miller (1985).  In that work they found
that the optical polarization spectrum of the archetypical type 2 Seyfert
galaxy, NGC 1068, could only be explained if our line of sight to its
nucleus were obscured, and a small fraction of the nuclear continuum were
reflected to us by electron scattering in warm, ionized gas concentrated along
the axis of the nuclear radio jet.  Miller \& Goodrich (1990), Tran, Miller, \&
Kay (1992), and Kay (1994) later found
similar behavior in many other type 2 Seyfert galaxies, and analogous regions
of warm electron scattering gas have been suggested in radio galaxies as well
(Barthel 1989; Antonucci \& Barvainis 1990; Antonucci, Hurt, \& Kinney 1994).

     On the basis of optical/ultraviolet spectropolarimetry and imaging, it
has been
possible to roughly estimate the physical conditions of this gas. In NGC 1068,
comparing the polarized line widths seen in a dust reflection nebula distant
from the nucleus to the polarized line widths seen in the nuclear reflection
region yields a temperature $\simeq 3 \times 10^5$K  and a Thomson optical
depth
$\tau_T \sim 0.01$ -- 0.1 (Antonucci \& Miller
1985; Miller, Goodrich, \& Mathews 1991).  {\it HST} imaging indicates a
characteristic dimension for the warm reflecting gas in this galaxy of
$\sim 10$ -- 100 pc (Kriss \etal 1993; Capetti \etal 1994).
Consequently, the ionization
parameter $\Xi$ (defined as $J_{ion}/(c p)$, where $p$ is the total gas
pressure) is $\simeq 25 L_{45}
[(\tau_T/0.01) (r/30\hbox{~pc})(T/3\times 10^5
\hbox{~K})]^{-1}$ (Krolik \& Begelman 1986).  Here we have
scaled to the luminosity estimated for NGC 1068 by Pier \etal (1994).
If the ionizing spectrum of NGC 1068 is similar to those (Figure 1)
of typical unobscured
AGN (as Pier \etal 1994 found), this ionization parameter and temperature
indicate (Figure 2) that in net, the gas absorbs more heat from the continuum
than it radiates, but it is not far from equilibrium (\cf the somewhat
idiosyncratic spectrum suggested by Marshall \etal 1993, for which this
gas might actually find itself in radiative equilibrium if the ionization
parameter lies within a special narrow range of values).  This placement
in the $(T,\Xi)$ plane suggests an origin for the gas in recent ionizing
evaporation from cooler gas, as the dynamics of this process tend to lock
the ionization parameter to values slightly in excess of critical (Krolik,
McKee, \& Tarter 1981; Krolik \& Begelman 1986).  Net radiative heating
drives the gas outward at mildly supersonic speeds.  That this description is
at
least qualitatively correct is supported by the fact that this picture predicts
(to within factors of a few) the correct temperature and optical depth of
the reflecting gas  (Krolik \& Begelman 1986; Balsara \& Krolik 1993).

    A few other observable signatures of this gas have been proposed, and,
in some cases, detected.  Krolik \& Kallman (1987) predicted that type
2 Seyfert galaxies with thick enough obscuration to block the 6 keV band
should display Fe K$\alpha$ lines of large ($\sim 1$ keV) equivalent width
and energies corresponding to rather high ionization stages.  These have,
indeed, been seen in several cases (Koyama \etal 1989; Marshall \etal 1993;
Awaki \etal 1990; Awaki \etal 1991; Ueno \etal 1994).
Band \etal (1990) showed that resonance scattering of the K$\alpha$ line
could also contribute to the observed feature.
Krolik \& Kallman ({\it op. cit.}) also predicted that type 1
Seyfert galaxies should show highly ionized Fe K photoionization edges with
optical depths $\sim 0.1$, and possibly OVIII ionization edges.

    It is the goal of this paper to predict a number of new observational
signatures of this gas, to make more quantitative the old predictions of
X-ray absorption features, and to compare these predictions to extant
observations.  In order of discussion, these signatures are:
the intrinsic emission spectrum; scattering; and absorption.  We concentrate on
the UV and X-ray bands because most of the intrinsic emission comes out in
the range 0.1 -- 10 keV and the strongest line features under the conditions of
interest are in the soft X-ray band.  In addition,
emission by stars in the host galaxy is relatively weaker in
these bands, and therefore easier to exclude.

   The work we present differs in a number of important respects from
earlier work on the X-ray irradiation of gas near Seyfert nuclei.  First,
the range of physical conditions we study is determined by dynamical models;
in some previous efforts (\eg Netzer 1993), the gas was assumed to
be strongly clumped, despite the absence of a confinement mechanism,
and the range of ionization states given the most attention is thermally
unstable for many common AGN spectral shapes.  Second, we find that
the scattering opacity due to resonance lines can be very significant,
and that a very large number of lines from many different ions contribute
(\S 4,5); in this context only Band \etal (1990) had previously recognized
this effect
of resonance lines, but they only considered the Fe K$\alpha$ line.  Third, in
all previous papers on this subject, the gas was assumed to be in ionization
balance (\eg Turner \etal 1993), and in many it was assumed to be in thermal
balance as well (\eg Yaqoob \& Warwick 1991; Netzer 1993; Fabian \etal 1994);
as we show here (\S 2,6) both of these assumptions are questionable.

\section{Calculational Method}

    Each of the properties we wish to predict depends on both $T$ and $\Xi$,
scaled by either an emission measure or the column density.  Our procedure
will be to use the photoionization code XSTAR (Kallman \& Krolik 1993) to
compute models at four representative points in $(T,\Xi)$ space: $(T = 3 \times
10^5~\hbox{K},~10^6~\hbox{K}) \times (\Xi = 10, 100)$, and scale the results
appropriately.  Given the estimates of the previous section, this box should
span most of the likely range of conditions for this gas.

    Details about the operation of XSTAR may be obtained, as is explained
in Kallman \& Krolik (1993), by anonymous ftp from legacy.gsfc.nasa.gov.
This code is particularly appropriate for this problem because of its
extensive X-ray line list, some 665 lines at energies above 120 eV.
In this application, we use XSTAR to compute the ionization balance in and
intrinsic emission from a succession of zones within a slab of fixed column
density.  These zones are chosen to be sufficiently thin that the continuum
radiation transfer is well-described, although this is not much of an
issue for our models because they are optically thin in most continuum
bands. Three passes through the slab
are made so as to converge on the correct line and continuum optical depths
from each zone to either side.  Emergent flux is corrected for any absorptive
opacity; when we cite intrinsic emission, we give the total emission through
both sides of the slab.  Total line and continuum optical depths
through the slab are a byproduct of this calculation.

    All our models are computed with the same spectrum, designed to be
representative of those seen in typical type 1 Seyfert galaxies and
low-luminosity quasars (see Figure 1).  We also consistently assume solar
abundances.  Since these lower luminosity AGN have
higher X-ray to optical luminosity ratios than the typical quasar, our ionizing
spectrum is considerably harder than that of Mathews \& Ferland (1987).
We chose our spectral shape to be the same as theirs at wavelengths longward of
2500 \AA, but at shorter wavelengths we use a power law of index 1.0
extending into the extreme UV.  This is a little flatter than slopes for
Seyfert 1's and 2's (e.g., Kinney et al. 1991), but corrects for steepening
due to Fe II and Balmer continuum emission, and it matches the low
luminosity QSO slope of 1.0 from O'Brien, Gondhalekar, \& Wilson (1988).
The observed hard X-ray spectrum of Seyferts is a sum of intrinsic emission
and a Compton-reflected hard tail.  We assume both components illuminate the
gas and use an energy index of 0.75 from 0.5 keV to 100 keV.
This matches the observed mean for Seyferts (Nandra \& Pounds 1994).
At 100 keV our spectrum breaks to an index of 2.4 to match the OSSE spectrum
of NGC 4151 (Maisack et al. 1993).
The X-ray normalization relative to the UV is determined assuming
$\alpha_{ox} = 1.25$, appropriate for Seyferts and low-luminosity quasars
(Kriss \& Canizares 1985).
The low energy X-ray spectrum steepens to an index of 2.0 below 0.5 keV.
Soft X-ray excesses in AGN span a large range of break point energies and
effective spectral indices, but our chosen values are near the middle of the
range observed by Walter et al. (1994).
This steep low energy X-ray spectrum meets the UV power law
at a breakpoint of 78.91 eV.

    Figure 2 shows the radiative equilibrium curve for this spectrum.  The
four points we have chosen span the region expected on the
basis of both empirical and theoretical arguments.  Examining models in
ionization equlibrium but out of radiative equilibrium is physically justified
on the basis of the following timescale ratios:
\beqar
{t_{recomb} \over t_{flow}} &\simeq & 5 \times 10^{-4} \left({\tau_T \over
0.01}\right)^{-1} T_{6}^{1.2} {\cal M} \\
{t_{heat} \over t_{flow}} &\simeq & 0.5  \left({\tau_T \over
0.01}\right)^{-1} T_{6}^{1.5} {\cal M},
\eneqar
where the recombination time has been evaluated for OVIII assuming that
the abundances of OVIII and OIX are equal, and the heating
time has used the net rate, \ie after subtracting off radiative cooling. This
latter point is significant because
the most likely regime is one in which the net heating minus cooling is
$\sim 0.1 \times$ the full heating rate.  For this reason, particularly
when the Mach number ${\cal M} \simeq few$,
adiabatic cooling can be significant, and the
flow time can be comparable to the time required for relaxation to
radiative equilibrium.

  The arguments of the preceding section demonstrate that ionization
balance should be a good approximation, despite the dynamical changes
in wind density and temperature, if the ionizing flux is constant.  However,
AGN are known to vary.  In \S 6 we will discuss the effects due to a variable
ionizing flux.

\section{Intrinsic Emission}

    All the processes by which gas in these conditions
radiates---bremsstrahlung, radiative recombination, and collisional
excitation---scale with density squared, so the most direct way of
describing the radiative output is in terms of an effective spectral
cooling function $\Lambda_\nu$ which is a function of $\Xi$ and $T$, and
an emission measure.  Note that in our regime, no strong line is affected
by collisional deexcitation, so there is negligible separate dependence on the
absolute density.

    The approximation of choosing a single value of $\Xi$ and $T$ amounts
to saying that the pressure scales simply as $r^{-2}$ everywhere and
the temperature is uniform.  Such simplicity is, of course, highly
unlikely, and is not found in hydrodynamic simulations (Balsara
\& Krolik 1993, 1994).  However, this approximation
may be adequate for our present purposes.

   Suppose that this gas expands outward with constant velocity and fixed
opening angle.  When that is the case, the density varies as
$n = n_o ( r / r_o )^{-2}$; thus the emission measure outside radius $r_o$ is
\beqar
EM &=& \int_{r_o}^{\infty} \, dr \, \Delta \Omega r^2 n_{o}^2 (r/r_o)^{-4}\\
 &=& \Delta \Omega n_{o}^2 r_{o}^3,
\eneqar
where $\Delta \Omega$ is the solid angle of its expansion cone.  Note that 3/4
of the total is accumulated within $2r_o$.  In the case of warm reflecting
gas in AGN, it is most useful to take $r_o$ as the radius at which the
gas expands out over the surrounding obscuration.  Gas closer in is, of
course, only partially visible when the nucleus itself is obscured.

Following Krolik \& Begelman (1986), we expect the temperature at $r_o$ to
be
\beq
kT_o \simeq (3\mu/5)^{1/3} \left({L \sigma_{eff} \over 3\pi r_o }\right)^{2/3}
\eneq
where the mean mass per particle is $\mu$, and $\sigma_{eff}$ is the absorption
cross section appropriately averaged over frequencies.  Because the heating
rate is $\propto r^{-2}$, while the adiabatic cooling rate is $\propto r^{-1}$,
if the velocity is constant we expect the temperature outside $r_o$ to
fall roughly as
\beq
T(r) = T_o {r_o \over r}\left[ 1 + \log (r/r_o)\right].
\eneq
By $r = 2r_o$, the temperature falls by only $\simeq 15\%$.  Similarly,
the ionization parameter (before allowance for continuum opacity) then varies
as
\beq
\Xi(r) = \Xi (r_o) {r/ r_o \over 1 + \log (r/r_o) }
\eneq
which likewise changes by only $\simeq 15\%$ from $r_o$ to $2r_o$.  Therefore,
most of the emission measure is accumulated under conditions of essentially
constant (unattenuated) $\Xi$ and $T$, and the approximation of taking
a single value for each of these quantities should be a reasonable one.

   The emission measure may also be rewritten in terms of quantities more
closely tied to observations.  It is
\beq
EM = C{\Delta \Omega L_{ion} \tau_T \over 4\pi c \sigma_T kT \Xi (n_e/n_H
+ n_i/n_H)},
\eneq
where $C \equiv \langle n^2 \rangle /\langle n \rangle^2$ is an allowance for
the possible effects of clumping.
Thus, if $T$ and $\Xi$ are relatively uniform over the entire sample of AGN,
the emission measure for any particular object, and hence the total inrinsic
X-ray flux, scales $\propto L \tau_T \Delta \Omega$.  On the other hand,
increasing $T$ and $\Xi$ decreases the emission measure.

\subsection{Soft X-rays}

   Results from XSTAR calculations are presented in Figs. 3.  All have
$\tau_T = 0.0665$ (corresponding to $N_e = 10^{23} \percmsq$).  The
normalizations assume that the irradiating continuum has a luminosity
of $10^{44}$ erg s$^{-1}$ in photons more energetic than 13.6 eV, and
that $C\Delta \Omega = \pi$.
Solid curves show the total emission from lines and continua smoothed by a
Gaussian of constant fractional width $\Delta \nu/\nu = 0.05$.
The dotted curves show the bremsstrahlung and recombination continua alone,
with no smoothing.

As is obvious from these figures,
the soft X-ray emission is very strongly line-dominated, even for the
higher values of $\Xi$.  OVIII Ly$\alpha$ is always strong, but a number of
other lines are often comparable.  We confirm the suggestion of Band \etal
(1990) that the Fe L recombination lines clustered near 1 keV are also strong.
If it were only for bremsstrahlung, the emission would be 1 -- 3 orders of
magnitude weaker, depending on the model and
the energy at which the comparison is made.  When the spectra are smoothed
by finite resolution, the lines blend into a pseudo-continuum, but the
strongest lines and line blends still stand out.  Note also that in a few
places the net emitted spectrum actually lies {\it below} the spectrum
due to continuum processes alone; this is an artifact of the smoothing
we imposed on the total spectrum but not on the pure continuum spectrum.

   Figures 3 also show that the emission measure scaling predicted by
equation 8 is borne out: in crude terms, $L_x \propto (T\Xi)^{-1}$.  However,
the complicated spectral shape means that the actual X-ray luminosity measured
by any particular instrument will depend strongly on the specific
energy-dependent sensitivity of that instrument.  Although
the emission is weak compared to the nuclear luminosity (by a factor of $\sim
10^{-3\pm 1}$), it is not necessarily negligible.  When our line of sight to
the nucleus is blocked, the emission from this gas should often be quite
noticeable (see \S 7.2).

   Several factors contribute to the weakness of the intrinsic emission:
the gas in these conditions is generally rather optically thin, so that
only a small fraction of the incident continuum energy heats the gas;
particularly in models c and d total heating exceeds total cooling; and
we normalize to a condition in which the gas covers only 1/4 of solid
angle around the source.

   In general terms, the spectrum above 1 keV can be expected to be quite
steep, but (as predicted by Krolik \& Kallman 1987), there is a very strong
ionized Fe K$\alpha$ emission line.  Below 1 keV, the spectrum is flatter,
particularly
for large $\Xi$ (models c and d).  The spectra predicted
by Netzer (1993) are significantly steeper than ours below 1 keV, probably
because his temperatures are significantly lower than any of those we
consider ($T = 1.3 \times 10^5$ K for his $U=3$ case).
Just as for the total luminosity, the
spectral slope measured by any particular instrument will depend strongly
on details of its energy-dependent sensitivity because the lines are so strong.
Therefore, to test a given model it is essential to fold the predictions
of that model through the response matrix for the experiment before making
any statements about whether it does or does not match the observed shape.

\subsection{Ultraviolet}

    Unless $C \gg 1$, the intrinsic emission of this gas in the ultraviolet
is very weak compared to the incident luminosity.  The continuum luminosity,
which is generated exclusively by bremsstrahlung, follows immediately from the
previously estimated emission measure:
\begin{equation}
{\nu L_{\nu} \over L_{ion}} = 2 \times 10^{-4} \nu_{15} C {\Delta \Omega
\over \pi} {\tau_T \over 0.0665} T_{6}^{-3/2} \Xi^{-1},
\end{equation}
where $\nu_{15}$ is the frequency in units of $10^{15}$ Hz.  The scaling
factors are likely to decrease this ratio still further.  Line emission,
particularly H and He recombination lines,
accounts for a larger luminosity.  For example, considering only the
H recombination lines, we find
\begin{equation}
{L(Ly\alpha) \over L_{ion}} = 3.6 \times 10^{-3}  C {\Delta \Omega \over \pi}
{\tau_T \over 0.0665} T_{6}^{-1.7} \Xi^{-1}.
\end{equation}
Other, collisionally excited lines, can also be strong, but their emissivities
depend in more complicated ways on $T$ and $\Xi$.

Although we expect intrinsic emission in the ultraviolet to be weak in
most instances, for the sake of completeness, and for use in the event of
large $C$, we present the actual spectral shapes in Figures 4.  In addition
to the intrinsic weakness of this emission, in practise it will be very
difficult to distinguish from narrow line emission (which should be much
stronger) unless an extremely small aperture is used.  Note also that the
Lyman edge emission jump decreases in strength as the temperature increases,
but should be easily observable up to temperatures close to $1 \times 10^6$K.

\section{Scattering}

   The contribution to the flux from reflection of the nuclear
continuum should always be comparable to or greater than the intrinsic
radiation.  To see this, consider the ratio of the two specific luminosities,
allowing only for the Thomson scattering portion of the reflection:
\beqar
{L_{\nu,intr} \over L_{\nu,refl}} &=& {C \Lambda_\nu (T,\Xi)
\Delta \Omega L_{ion} \tau_T /( 4\pi c \sigma_T kT \Xi) \over (\Delta
\Omega/4\pi) \tau_T  L_{\nu,nucl}}\\
 &=& {C\nu \Lambda_\nu (T,\Xi) \over c kT \sigma_T \Xi}{L_{ion} \over \nu
L_{\nu,nucl}}\\
 &\simeq & {8.3 C\over \Xi T_6} {\nu \Lambda_\nu \over 10^{-23} \hbox{erg
cm$^3$ s$^{-1}$}} {L_{ion} \over \nu L_{\nu,nucl}}.
\eneqar
In all four of the models considered here, $\nu \Lambda_\nu$ in the soft
X-ray band is in fact within a factor of two of $10^{-23}$ erg cm$^3$ s$^{-1}$.
For $T_6 \ltwid 1$ and $\Xi \sim 10$ -- 100, it is clear that the intrinsic
soft X-ray emission and reflection by Thomson scattering are comparable.
In the ultraviolet, reflection should completely dominate intrinsic emission
(unless the clumping factor is quite large) because most of the cooling
radiation from gas in this state emerges in the soft X-ray band.

  Thomson scattering is not the only sort of scattering significant in
this situation, although that assumption has frequently been made in
cognate work (\eg Netzer 1993).  As first suggested for the Fe K$\alpha$
line alone by
Band \etal (1990), resonance line scattering can substantially enhance the
scattering opacity.  Our calculations show that so many lines have the
potential to be optically thick in these conditions that taken together they
can
increase the frequency-averaged scattering depth by factors of a few.  Details
such as exactly which lines participate, and how large their optical depths
are,
obviously vary with $T$ and $\Xi$.

    However, there is another
important consideration which renders any quantitative prediction of resonance
line scattering rather model-dependent: the strong velocity gradients which
are likely to exist.  If this warm gas is a pressure-driven wind, as suggested
by Krolik \& Begelman (1986), then we can expect its mean speed in the
region we see to be a few times the sound speed.  We can also expect that
the geometrical divergence of the wind convolved with velocity gradients
parallel to streamlines will introduce velocity differences across the wind
of magnitude comparable to the typical wind speed.  Because X-ray lines all
arise in elements with $Z \sim O(10)$, the local thermal width of atomic
speeds is at least an order of magnitude smaller than the actual
velocity differences across the wind.  Therefore, the equivalent width
of an optically thick line is very sensitive to the velocity differences
across the wind, which we do not know to anything better than order of
magnitude accuracy.

   In order to illustrate in at least a tentative way how lines may alter
the reflecting properties of this gas, we have computed the reflected fraction
under the assumption that the true velocity distribution of the gas is
$ \propto \exp[-(v/c_s)^2]$, where $c_s$ is the sound speed.  While probably
not literally correct for any plausible dynamical model, this
form for the distribution does at least possess the desirable qualitative
properties of being peaked at the center and having a width comparable to the
sound speed.  We also assume, as we
did for our prediction of the intrinsic emission, that the covering fraction
is 1/4.

  As discussed in \S 5, there are wavelengths for which this gas
can have modest continuum optical depth.  We account for this in the reflected
fraction by making the approximation that the mean pathlength for reflected
photons is equal to the pathlength for photons travelling straight outward
through the gas, and multiplying by the appropriate $\exp(-\tau_\nu)$.
In more weakly ionized gas, continuum opacity has a stronger effect
on the reflected spectrum (Netzer 1993); here, however, it has only minor
impact.

   The results for the soft X-ray band are
shown in Figures 5.  While there are line-free portions of this band,
it is clear that on average resonance scattering can increase the reflected
fraction by factors of a few relative to the Thomson-only reflected
fraction (which for these numbers is 0.0166).  The energy range from 1 -- 1.5
keV is particularly rich in such lines.  As predicted by
Band \etal (1990), resonance scattering does significantly enhance the
strength of Fe K$\alpha$ in the reflected spectrum, but this is also true
of many other lines, and, as we have already emphasized, the amount of
enhancement depends very strongly on the velocity structure of the gas.
It is also important
to observe that when the gas is relatively weakly ionized (Model b),
absorption can actually cause the reflected fraction to fall below the
Thomson value.

   Figures 6, in which we combine reflected and intrinsic emission from 0.1
to 10 keV, show the
actual spectrum that might be observed when the nucleus itself is obscured.
Both the intrinsic and reflected contributions scale $\propto \Delta \Omega
\tau_T$; as before, we take $C\Delta \Omega = \pi$, and $\tau_T = 0.0665$.
The addition of reflected nuclear continuum to the intrinsic
emission has several consequences for the observed spectrum.  First, of course,
the total flux is increased by a factor of a few. Second, the
equivalent widths of emission lines which are either forbidden or arise from
excited states are diluted, while the equivalent widths of permitted resonance
lines are enhanced.   Third, the spectral shape is altered, but in a sense
which could vary substantially from object to object because Type 1 Seyfert
galaxies, in which we can see the nuclear soft X-ray flux unobscured, show
a wide dispersion in soft X-ray spectral slopes
(Turner \& Pounds 1989; Walter \& Fink 1993; Walter et al. 1994).
For the purposes of this example, we use the same shape spectrum (Figure 1)
that we used for the photoionization calculations.

    Although details of the reflected plus intrinsic spectrum depend on
the choice of $\Xi$ and $T$, one feature is comparatively stable: the
large equivalent width Fe K$\alpha$ line (Krolik \& Kallman 1987;
Band \etal 1990).  As noted by Band \etal (1990) this feature can contain a
substantial contribution from resonance scattering in addition to the
intrinsic emission.
Depending on $T$ and $\Xi$, the reflected luminosity in the Fe K$\alpha$
feature relative to the intrinsic emission of the gas in our models
varies from roughly a tenth to of order unity.
Another prominent feature is the Ly$\alpha$ transition of OVIII.
This depends rather more on physical conditions,
but it remains strong over most of the expected range.

    Before leaving the subject of the observed X-ray spectrum in obscured
cases, it is also important to point out that the observed range of column
density in the obscuring material is fairly broad: from as little as
$\sim \rm few \times 10^{22}~\percmsq$ up to $ > 10^{24}~\percmsq$
(Mulchaey \etal 1993).  When the Thomson optical depth is at least
a few, the nuclear flux is sufficiently suppressed that the observed
spectrum (below 10 keV) should be fairly close to the examples shown in Figures
6.  There is a small ``Compton reflection" contribution from X-rays
scattered off the inner edge of the obscuration, but this is weak except
in rather special geometries and for viewing angles nearly out of the
obscured range (Krolik \etal 1994; Ghisellini \etal 1994).

   However, when the column density is
smaller, the nuclear flux can shine through at energies above several
keV.  In that case, Figures 6 should give a good representation of the
X-ray spectrum at lower energies, but above the point where the obscuration
becomes optically thin (somewhere between a few and 10 keV), the spectrum
will rise steeply and then resume the usual power-law behavior at
higher energies (Krolik \etal 1994; Ghisellini \etal 1994).  This steep
rise also includes a contribution from ``Compton reflection" off the
inner edge of the obscuration, but its magnitude is, of course, small
when the Thomson optical depth of the obscuration is small.

In the ultraviolet, very few lines are optically thick even in the limit of
zero velocity gradient.  By far the most
important are the H Lyman series and the resonance doublets of the Li
isoelectronic sequence (see Table 1).  Ly$\alpha$ has a significant optical
depth in all four models, but only in the most weakly ionized model
(Model b) do any others become optically thick, and then it is only the
two lines NV $\lambda$1240 and OVI $\lambda$1034.
Nonetheless, resonance lines can produce
emission features in the reflection spectrum even when optically thin
if their optical depths are greater than the continuum optical depth.
In favorable circumstances, therefore, the equivalent width of Ly$\alpha$,
or possibly OVI $\lambda$1034, in the reflection spectrum might be as large as
$\sim \lambda (v/c)/\tau_T \sim 100\AA$.  The other lines should,
in general, be rather weaker.

Detection of these weak UV reflection features is made even more difficult
by the fact that
they are superposed on other emission lines.  In total flux spectra (unless
extremely small apertures are used) they
are hidden by narrow line emission; in polarized flux spectra---which
eliminate the narrow lines---they are blended with the reflected broad emission
lines.

\section{Absorption}

  The transmission fraction follows directly from these same models.  It is
simply
\beq
f_\nu = \exp[-\tau_{c}(\nu) + \tau_l (\nu)] + f_{refl}(\nu)
\eneq
where $\tau_{c}(\nu)$ is the continuum optical depth at frequency $\nu$ and
$f_{refl}(\nu)$ is the reflected fraction at that frequency.  We
compute the line optical depth $\tau_l (\nu)$ in the same way we did
for the reflected fraction, {\it i.e.} we assume that all the gas has a
Gaussian
velocity distribution function with characteristic width equal to the
sound speed, and sum over all lines with frequencies sufficiently near
$\nu$.

   There is one problematic element in evaluating $f_{\nu}$: the
expression just given assumes that
the total column density through which the continuum passes is the
same as the column density visible by reflection.  While this is certainly
the minimum, there may be additional matter in regions obscured by the
torus.  We also know much less about the obscured gas's physical
state.  Our only guide is hydrodynamical simulations, such as those
reported in Balsara \& Krolik (1993, 1994).  Unfortunately, these simulations
indicate that whether the reflection region can be used to make a good
prediction of $\Xi$ and $T$ in the hidden region depends on additional
poorly known parameters such as the detailed shape of the inner edge
of the torus and its specific angular momentum.
To cope with this uncertainty, the results we present (Figures 7) show the
effect of only the visible gas; in order to estimate the impact of the
hidden gas the best one can do is imagine multiplying the optical depths
by factors of several.

In all four models we have examined, three features stand out most clearly
in the predicted transmission spectra: K edges of OVII or OVIII; the K
edges of the Fe ionization stages within a few of FeXX; and the L edges
of those same Fe ionization stages.  Their relative strengths vary, of
course, depending on the parameters, but in most of the likely range
the absorption trough near 1 keV produced by OVII/OVIII and Fe L-absorption
(plus assorted lines) is larger than the Fe K absorption.  Only in
the most highly ionized case (Model c) is the Fe K edge competitive
with the 1 keV feature.

Calculations of soft X-ray opacity have traditionally included only
photoelectric absorption and sometimes Compton scattering (\eg Krolik \&
Kallman 1984; Netzer 1993; Turner \etal 1993).  We find that
in these circumstances ({\it i.e.} relatively strong ionization), while
photoelectric absorption is the most important source
of opacity, resonance line scattering can also be significant.
While our transmitted spectrum for Model b is qualitatitvely similar to
Netzer's (1993) U=10 case (the closest to ours in ionization level), there
are important differences more apparent in the higher ionization models.
For example, in Model a the optical
depth due to merged resonance lines in the vicinity of 1 keV is typically
$\sim 0.3 \times$ the optical depth due to photoelectric absorption,
although of course this figure varies by factors of several as a function
of wavelength, and depends significantly on the degree of velocity
broadening.

  The placement of line opacity can also be important.  Again
referring to Model a, we find that the apparent low energy edge of the
absorption trough due to K-shell photoionization of O falls at $\simeq 750$ eV,
nearly equal to the OVII edge of 739 eV,
even though the abundance of OVIII, whose ionization threshold is 871 eV,
is $\sim 100 \times$ greater than the abundance of OVII in this
model.  Thus, merged lines could easily cause a misidentification of
the O ionization state.

   To be precise, the total spectrum one measures is the sum
of the direct flux as filtered by absorption (Figures 5) and the intrinsic
emission (Figures 3).
This same point is emphasized by Netzer (1993).
Our definition of absorption already takes into account
scattering.  However, unless the optical depth is great enough to absorb
a significant fraction of the continuum power, the luminosity of the
intrinsic emission is a small fraction of the direct luminosity; for example,
in model a even the OVIII L$\alpha$ line has an equivalent width relative
to the transmitted continuum of only $\simeq 10$ eV.  We expect,
therefore, that even the strongest intrinsic lines will only occasionally
have enough equivalent width to be visible.

In the ultraviolet, the situation is quite different.  As remarked above,
only a few lines (the H Lyman series, OVI $\lambda$1034, and NV $\lambda$1240)
have any significant optical depth, and the last two only become interesting
if the ionization level is as low as in Model b.
The only continuum feature, of course, is the Lyman edge, but its optical
depth in these conditions is generally $\sim 0.1\tau_T \ll 1$.  Consequently,
we expect this gas to have little impact on the UV spectrum seen in
transmission.

  The absorption and scattering we predict might, in some circumstances,
generate a significant radiation force on the gas (\cf Reynolds \& Fabian
1994).
In most cases ({\it e.g.}
Model a), the increase above the radiation force due to Thomson scattering
is only a few tens of percent.  Although the opacity at 1 keV is several
times Thomson, the band in which the opacity is enhanced carries only
$\sim 10\%$ of the total flux.  However, if the gas is more weakly
ionized (as in Model b), the radiation force might be as much as double
that due to Thomson scattering alone.

\section{Time-Dependent Effects}

   In our discussion so far we have assumed that the ionizing flux is
constant.  In view of the substantial variations often observed in the UV
and X-ray fluxes of AGN, it is now time to examine that assumption
more carefully.

Consider a model equation for the abundance of a particular ionization stage
of a photoionized element:
\begin{equation}
{d n_i \over dt} = - [F_i \sigma_{ion,i} + n_e \alpha_{rec,i-1}] n_i +
n_e n_{i+1}\alpha_{rec,i} + F_{i-1} \sigma_{ion,i-1} n_{i-1},
\end{equation}
where we have neglected Auger ionization, collisional ionization, and
three-body recombination.  The products $F_i \sigma_{ion,i}$ are meant
to be the ionizing (photon number) fluxes for stage $i$ appropriately
integrated over
the frequency-dependent ionization cross sections $\sigma_{ion,i}$, while
$\alpha_{rec,i}$ is the recombination coefficient for producing stage $i$.
If the ionizing flux is a function of time, the (formal) solution of this
equation is
\begin{equation}
n_i (t) =  \exp[-\int_0^{t} \, dt^{\prime} \, P_i(t^{\prime})]
\int_0^t \, dt^{\prime} \, [n_e n_{i+1} \alpha_{rec,i} +
F_{i-1}\sigma_{ion,i-1} n_{i-1}] \exp[\int_0^{t^{\prime}} \, dt^{\prime\prime}
\, P(t^{\prime\prime})],
\end{equation}
where
\begin{equation}
P_i (t) = F_i (t)\sigma_{ion,i} + n_e \alpha_{rec,i-1}.
\end{equation}
That is, the abundance of stage $i$ is given by the integrated creation
rate of stage $i$ extended over a time $t_d$ which is roughly the typical
time for destruction of an ion of stage $i$, whether by ionization or by
recombination.

   The character of the time-variation of ionic abundances depends on
two ratios: $t_d/t_{var}$, where $t_{var}$ is the shortest timescale on which
order unity fluctuations in the ionizing luminosity occur, or, if the
fluctuations are never that great, the timescale on which
most of the variance in the ionizing luminosity is accumulated; and
$\langle (L_{ion} - \langle L_{ion}\rangle)^2 \rangle^{1/2}/\langle L_{ion}
\rangle$, the ratio of the {\it rms} variation amplitude to the mean
value of the ionizing luminosity.  When $t_d \ll t_{var}$, the
ionization state is in equilibrium with the ionizing flux, but there
is no steady-state.  However, if the amplitude of variability is small,
the departures from the mean are small.  When $t_d \gg t_{var}$,
the ionic abundances reach a steady state defined by the mean values of the
ionization and recombination rates, but this steady state is almost
never in equilibrium with the instantaneous value of the ionizing flux.
Again, the differences are small unless the amplitude of variation is
substantial.  Finally, if $t_d \sim t_{var}$, the ionic abundances follow
a history which is a smoothed and delayed version of the history of the
ionizing flux.  In this case, the abundances are neither in a steady state
nor in equilibrium.  No matter what the value of $t_d/t_{var}$ is, the
abundance of ion $i$ cannot change substantially in a time short compared
to $t_d$; in fact, observed variations in the relative abundances may
be used to place an upper bound on $t_d$, and hence an upper bound on
the distance between the absorbing matter and the source of ionizing
photons.

  To decide which case applies, we must estimate $t_d/t_{var}$.  Before
doing so, we point out that $t_d$ can vary by orders of magnitude from
one ion to the next, even for ions of the same element.  Typical continuum
spectra can often be described by $F_\nu \propto \nu^{-\alpha}$ with
$0.5 \ltwid \alpha \ltwid 2$.  Ionization potentials in mid-Z elements
range over roughly two orders of magnitude in energy, so the number fluxes
of ionizing photons can have mutual ratios as large as $\sim 10^{2\alpha}$.
Consequently, different ionization stages of the same element can be in
quite different regimes with respect to the ratio $t_d/t_{var}$.

Earlier, when we argued that, measured on a flow time, the ionization balance
stays very close to equilibrium, we estimated $t_d$ (for the fiducial ion
OVIII) by setting it equal to $t_{recomb} \equiv [n_e \alpha_{rec,i}
n_{i+1}/n_i]^{-1}$; that is appropriate when $n_{i-1} \ll n_i$ and
the balance is near equilibrium.  Using the same approximation, but
evaluating $t_{recomb}$ for OVIII in time units, we find
\begin{equation}
t_{recomb}(OVIII) = 12 {n(O VIII)
\over n(O IX)} \left( {\tau_T \over
0.01}\right)^{-2} T_{6}^{-0.3} \left({\Xi \over 10}\right)^{-1}
L_{ion,44}~\hbox{yr},
\end{equation}
where $L_{ion,44}$
is the ionizing luminosity normalized to $10^{44}~\ergs$.  Note that
the abundance ratio depends strongly on conditions: in our models it ranged
from $0.003$ in Model c, to $\simeq 0.03$ in Models a and d,
to $\simeq 0.9$ in Model b.

 There is no object in which $t_{var}$ is well determined.  The
timescale on which most of the variance is accumulated seems in many cases
to be $\gtwid 1$ yr, and the amplitudes of the fluctuations are generally
order unity compared to the very long-term time average fluxes (Edelson,
Krolik,
\& Pike 1990 for the UV; Green, McHardy, \& Lehto 1993 for the X-ray).
In many cases the power spectrum of fluctuations has a power-law character
over a substantial dynamic range in frequency, so there can also be
fluctuations of order unity compared to the local mean flux on much shorter
timescales.
Given this wide range of variability properties, our estimate for
$t_d ($OVIII) indicates that $t_d$ for that ion
could be anywhere within one to two orders of magnitude larger or
smaller than $t_{var}$.  Consequently, one must be very careful about
which mean flux level is used to compute the current ionization state.

   The properties we observe are further averaged over the region.  True
emission and scattering are integrated over the entire volume, requiring
an average over the light crossing time, possibly years to decades or even
centuries.  If
$t_{var}$ is at most a few years, this average may
be comparatively stable.  However, absorption carries
information only about locations on the direct line of sight, so all
the sub-regions may be regarded as effectively coherent in time.  Thus,
absorption properties are directly sensitive to the problems of ionization
disequilibrium we have just pointed out.  This provides yet an additional
reason why ``predicted" transmission spectra such as those shown in Figure 7
must be regarded as illustrative at best.

\section{Comparison with Observations}

\subsection{Unobscured AGN}

   {\it Ginga} and {\it Rosat} data indicate that the X-rays in a number of
unobscured AGN pass through ``warm absorbers" on their way to us (\eg Pounds
\etal 1993; Nandra \& Pounds 1994).  The main observational signatures of this
material are Fe K edges with optical depths $\sim 0.1$ at energies
corresponding
to ionization stages of Fe around FeXX, and edges of OVII and OVIII with
optical
depths $\sim 0.1$ -- 1.  ASCA data, with its much greater spectral resolution,
should permit a tremendous refinement of this picture (Inoue 1993).
Because these
identifications have been found by the model-fitting which is customary
in X-ray spectroscopy, it is quite possible that many other features may also
be present.  Absorption of this character is very nicely in line with
the predictions we have made here.  We therefore suggest that these ``warm
absorbers" can be identified with the warm, reflecting gas we have been
discussing in this paper (\cf Krolik \& Kallman 1987, Nandra \& Pounds 1994),
and its inner, hidden extension.

   Interpreting these data requires great care.  Both true absorption
opacity and scattering must be included if one desires a reasonably
accurate description of the frequency-dependent opacity.  Blended lines
may significantly shift the apparent energy of absorption edges, leading
to misidentification of the ion responsible.  Unfortunately,
as we discussed in \S 4, consideration of resonance line opacity requires
the introduction of at least two more free parameters, one specifying
the covering factor, the other the width of the material's velocity
distribution.

   In addition (a similar point was made by Netzer 1993 in a somewhat different
context), because a large number of lines and edges contribute to the
absorption opacity, if one fits models to the data it is preferable to
fit well-defined complete photoionization models, rather
than individual features.  This procedure does not materially increase
the number of free parameters (one need specify only $T$, $\Xi$, and $\tau_T$,
as compared to, \eg the energies and optical depths of several edges), but
it should improve the quality of the fits, and possibly reveal features that
might otherwise be left as residuals.

   However, even this procedure has
its limitations.  First, contrary to what is frequently done, in
this situation {\it there is no need to impose thermal balance} on the
models (\S 2) (in fact, the models of Turner \etal 1993 are an example of
this practice---their temperature was fixed, somewhat arbitrarily, at
$10^5$K).  Rather than finding that ionization parameter whose equilibrium
temperature is such that the ionization structure best fits the data,
one can only restrict the
state of the warm absorber to a region in the $\Xi$ -- $T$
plane.  For example, on the basis of the models shown here, we can
already say that in most type 1 Seyfert galaxies, the column density of
material as weakly ionized as in Model b must be considerably less than
$10^{23}$ cm$^{-2}$; otherwise there would be consistently more soft
X-ray absorption observed.

    Second, there are likely to be numerous examples
in which {\it there is no need to impose strict ionization balance} on
the models (\S 6). Whether ionization balance can be expected in any
particular case must be evaluated on the basis of the specifics of that
case---which ion(s) are important, the mean luminosity of the AGN, and
the variability properties of that AGN.  The general sense of the effect
of departures from ionization balance is that the ionic abundances we observe
reflect an average over the recent history of the AGN's ionizing luminosity,
and should therefore change more slowly than the instantaneous luminosity.
Additional complications are introduced by the fact that different ions
average over different lookback times.
However, even this generality cannot be followed blindly, for most AGN
light curves are drastically under-sampled.  Consequently, it can
be very difficult to estimate the correct average fluxes, or even
determine whether substantial variation in the ionizing flux has occurred.

   The behavior of the ``warm absorber" in MCG 6-30-15 observed by
Fabian \etal (1994) may be an example of several of the subtleties
discussed in the preceding paragraphs.  They saw the
optical depth of a feature which they identified as the OVII K edge change
from $\simeq 0.6$ to $\simeq 1.2$ in the span of a month, while the
X-ray flux dropped by about a factor of 2.  Using photoionization models
requiring both thermal and ionization equilibrium, they inferred that
the column density had increased while the ionization parameter was
roughly constant.  We suggest several possible reinterpretations of
these data.  As we pointed out earlier (\S 5), a merger of
line features with an OVIII edge can masquerade as an edge at the
energy of OVII.  This identification gains some support from the feature in
the residuals spectrum at the energy of the Fe L edge.  If this suggestion
is correct, the observed change in optical depth would be consistent
with fixed column density and ionization equilibrium because in the
regime where OVIII dominates OVII, the OVIII abundance itself is
inversely proportional to the X-ray flux.  However, given
the ionizing luminosity of this AGN ($\sim4 \times 10^{43}~\ergs$, from
data in Walter \& Fink 1993),
the expected recombination timescale for OVIII is $\sim 40 (\tau_T/0.01)^{-2}$d
if, for example, the conditions of Model a obtain.  That would mean the
ionization states seen in the two measurements were controlled by the
ionizing flux averaged over roughly that length of time preceding the
two observations.  Because the 1 -- 10 keV flux in that object has been
seen to vary over a range of a factor of three in as short a time as
three days (McHardy 1990), the instantaneous fluxes measured by {\it ASCA}
need not be very good guides to the appropriate average fluxes.

    Finally, we point out that most of the observed narrow absorption lines
seen in the UV spectra of a number of AGN
(Voit, Shull, \& Begelman; Bahcall \etal 1993; Aldcroft \etal 1994)
are almost certainly {\it not} due to this warm gas. The
warm gas we discuss here might produce measurable absorption in
Ly$\alpha$ and OVI $\lambda$1034 with an
expected width and offset to the blue of several hundred \kms.
In fact, Ly$\alpha$ absorption blueshifted by $\sim1000$ \kms with no other
corresponding metal lines {\it is} seen in the spectrum of NGC~5548
(Korista \etal 1994).
CIV absorption, however, must come from some other material.  In all four of
our models, the CIV optical depth (before allowance for velocity gradients)
was never more than $\sim 10^{-2}$ because in all cases, the overwhelming
majority of C atoms were completely stripped.  If there were to be any
significant abundance of CIV, the mean ionization stage of Fe would be
far below the level inferred from the positions of the observed Fe K edges.

On the other hand, OVII and OVIII edges of moderate optical depth can
be produced in either of two ways: They can be created by the conditions
discussed in this paper, in which the column density is relatively large,
and most O is completely stripped; or they may be created by gas more
weakly ionized (so that most O is either OVII or OVIII) having rather
smaller column density ($10^{21}$ -- $10^{22}$ cm$^{-2}$).  Those modelling
efforts which have imposed both ionization and thermal balance have tended
to home in on the latter solution because it is achieved for $\Xi$
just below the critical value beyond which no cool equilibria exist.  In this
state, while most C and N atoms are likewise stripped down to
the K shell, there are enough Li-like ions of C, N, and O to produce
significant opacity in the resonance doublets  CIV  $\lambda$1549,
NV $\lambda$1240, and OVI $\lambda$1034
(Mathur \etal 1994).  However, if the gas is to possess any significant
opacity in the transitions of still more weakly ionized species (\eg
the MgII $\lambda$2798 line: Mathur 1994),
the abundance of OVII and OVIII would certainly be negligible.

These two different mechanisms for creating the observed OVII and OVIII
edges may be distinguished most easily by searching for Fe absorption
edges.  To produce either an Fe K or an Fe L edge of optical depth
$\sim 0.1$ or greater requires a column density of at least $\sim 10^{23}$
cm$^{-2}$.  Thus, Fe K features of the sort seen in {\it Ginga}
spectra of AGN (Nandra \& Pounds 1994) require the large column density,
high ionization state solution; a significant Fe L edge would require
the same interpretation.  On the other hand, the absence of these features
would favor the small column density, lower ionization state solution.

It is possible that material in both states exists in individual AGN.
As material is photoionized, heated, and driven off the inner
surface of the obscuring torus, it must pass through intermediate stages
of ionization before reaching the quasi-steady state on which we have
focussed here.  In these intermediate stages it would be capable of
imposing enough UV line opacity to create the observed UV absorption lines;
it would also possess significant opacity in the OVII and OVIII K edges.
However, the time-averaged state of the gas would be more highly ionized,
so at any one time most of the gas mass would be in the more highly
ionized state.  The two states would then produce comparable amounts
of OVII/OVIII K edge opacity, but the more highly ionized gas would
be responsible for all of the ionized Fe K and L edge opacity.  When this
condition exists, interpretation of time-variable O photoionization
edges becomes even
more complicated than indicated in our discussion of MCG 6-30-15.

\subsection{Obscured AGN}

    As we have already remarked, the primary emission observable in the UV
from this gas is the nuclear reflection.  This is due mostly to
Thomson scattering, which has been amply discussed in the literature,
but those few UV resonance lines whose optical depths are greater
than the electron scattering optical depth will appear as emission
features in the reflection spectrum.  Because the nuclear spectrum
generally shows broad emission features at these same lines, the
reflection contribution to the feature will appear as a comparatively
narrow component
superposed on the broad emission line profile. Depending on the
projection of the velocity of outflow on our line of sight, this
narrow component could be either redshifted or blueshifted by as
much as a few hundred km $s^{-1}$.  Note that broad emission lines
which are not resonance transitions ({\it e.g.} the Balmer lines
or CIII] 1909) should {\it not} possess narrow components due to
reflection.

In the X-ray range from 0.1 -- 10 keV,
the warm gas should be visible {\it via} a mixture of
Thomson reflection, resonance line reflection, and intrinsic emission.
The resulting X-ray spectrum should be very complex.

   Only for NGC 1068 are there published data of sufficiently high
quality to make a comparison with these predictions (Marshall \etal 1993;
Ueno \etal 1994).  In this case there is a large excess below 2 keV relative
to the extrapolated high energy power law, but about half of
this flux is due
to extra-nuclear sources (Wilson \etal 1992).  Many emission lines are clearly
present between 0.5 and 2 keV, but their identifications are somewhat
uncertain, and there is no clear way to distinguish between a nuclear and
an extra-nuclear origin for these lines.  Nonetheless, the fact that there
is as much unresolved flux in this band as there is indicates that the
gas in the reflection region must be more highly ionized than Model b
because otherwise very little in the way of soft X-rays would emerge.

In some cases of unobscured AGN (\eg IC 4329A: Madejski \etal 1994) the
column density of the warm absorber appears to be rather small,
$\sim 10^{21}~\percmsq$.
These objects viewed from the side would have much weaker
reflection and intrinsic emission, for both scale $\propto \tau_T$.  All
that we would see in X-rays from these galaxies would be whatever portion
of the direct nuclear flux can penetrate the obscuration (see Krolik \etal
1994 or Ghisellini \etal 1994 for explicit calculations of how large this
might be).  Those type 2 Seyfert galaxies with little polarization (Kay 1994)
may be objects like these viewed from an obscured direction.

\subsection{NGC 4151}

The case of NGC 4151 is a bit unusual since it combines features of both
the obscured and the unobscured AGN.  The kilovolt X-ray continuum is heavily
absorbed, but appears to not be completely obscured.
The optical and UV continuum is directly visible, yet the spectrum is rich
in absorption lines.
The most popular model for the X-ray absorption has been a large column
($\sim 1 \times 10^{23}~\percmsq$) of cold gas which only partially covers
the source of the X-ray continuum (\eg Holt \etal 1980).
Recently, however, Weaver \etal (1994a, 1994b) have shown that a warm
absorber model that includes partial reflection of the nuclear continuum
can fit the X-ray spectrum.
The soft X-ray flux in these fits
is almost entirely due to a reflection component,
leaving little room for intrinsic emission.
The historical lack of variability in this soft component
(Perola \etal 1986; Pounds \etal 1986; Weaver \etal 1994a)
suggests reflection from a source distributed over a region
light years in diameter.

We identify the agent of reflection
with the warm reflecting gas we have discussed here.
If this identification is correct, we would expect the
soft X-ray spectrum of this source to exhibit the merged line features shown
in Figs. 6.
The reflected fraction in NGC~4151
($\simeq 2.5\%$: Weaver et al. 1994b) implies $\tau_T = 0.1$ for
$\Delta \Omega / 4 \pi = 0.25$.  This leads to a predicted
size $\simeq 0.05 (\Xi/10)^{-1} T_6^{-1}$ pc, which would be consistent with
the lack of variability if $T_6 \Xi$ were as small as $\simeq 1$ -- 2.

The morphology of the ionized gas imaged with {\it HST} implies
that our line of sight passes through a substantial column of optically thick
material (Evans \etal 1993).  In the context of the obscuration/reflection
model, this fact suggests that the warm material on our line of sight is
likely to be in a lower state of ionization than the bulk of the warm
reflecting gas.
The Fe K edge energy also indicates an ionization level similar
to, or lower than, our Model b.  For this model, the timescale for changes
in soft X-ray opacity is $\simeq 12 ( \tau_T / 0.1 )^{-2}$ d,
where we have used an ionizing luminosity of $2.1 \times 10^{43}$ erg s$^{-1}$
(Evans et al. 1993) and the optical depth derived above.
(Note that this timescale is significantly longer than the 100 s used by
Weaver \etal 1994a, who make the tacit assumption that the absorbing gas
is similar to broad-line clouds in density and therefore located much
closer to the source for similar ionization parameters.)
Our 12 day timescale is consistent with both the BBXRT (Weaver \etal 1994a)
and the ASCA (Weaver \etal 1994b) observations.
The BBXRT data show a possible, slight increase in
absorption below 2 keV in just 0.7d,
while the 2 -- 10 keV flux decreases by about $40\%$.
Given the smoothing effects of the 12 d recombination timescale, one expects
only small changes in opacity over a day in response to even large
continuum fluctuations.
The ASCA data (Weaver \etal 1994b), comprising two observations separated by
about five months, show large changes in soft X-ray opacity, even while
the intrinsic 2 -- 10 keV flux increases by only about 20\%.
This behavior is again consistent with our previous arguments that the
ionization structure need not reflect the instantaneous continuum, but
rather an average over the variations occuring over roughly the preceding
recombination timescale.  This would require monitoring for $\sim12$ days
to adequately constrain a physical model for the absorber.

\section{Conclusions}

We have shown that X-ray heated winds in AGN may manifest their presence
in many ways.  In addition to serving as the warm reflecting gas that
permits us to view the inner regions of obscured AGN, the wind may
produce the ``warm" absorption seen in many AGN X-ray spectra.
Our models provide a qualitative match to many of the observed properties
of both obscured and unobscured AGN.

The X-ray emission from the wind is dominated by line radiation superposed
on bremsstrahlung plus recombination continua.
This may be visible in obscured AGN if the nucleus can be isolated from the
surrounding galaxy, but the nuclear continuum reflected by Thomson scattering
and by resonance line scattering will be comparable or greater in brightness.
The strong kilovolt emission lines visible in the X-ray spectra of obscured
AGN may be produced by both intrinsic emission and resonance
scattering in the X-ray heated wind.
The UV emission is principally bremsstrahlung
plus recombination lines of H and He.
Reflection of the nuclear continuum by Thomson scattering will dominate the
UV spectrum, but there will be some contribution from resonance
lines.  These lines would be difficult to distinguish from emission in the
narrow-line region, but they might be most easily visible as narrow features
in a polarized flux spectrum.

In unobscured AGN the reflected and intrinsic components of the gas are
swamped by the nuclear continuum, but absorption by the resonance lines and
photoionization edges of highly ionized species will imprint their signature
on the transmitted spectrum.  The presence of merged line features can
seriously complicate
the identification of edges.  The high columns required to produce Fe K
edges in {\it Ginga} X-ray spectra and comparable OVII and O VIII edges
in {\it ROSAT} spectra correspond nicely to our predictions.

Absorption in the UV, however, is expected to be weak.
Only the Lyman lines reach significant optical depth, and there is some
contribution from the OVI and NV resonance doublets in the lowest ionization
models.  Most UV lines observed in AGN must come from lower ionization gas
with a lower column density than that producing the Fe K edges; this
gas could be an intermediate state in the production of the more highly
ionized reflecting gas.

Time scales for variations in the emitted and reflected components of the
wind should be years or longer since the gas fills a region parsecs
or more in size in a typical AGN.  Sight lines for absorption, however,
are effectively coherent in time, and the time-dependent effects are a
function of the variability timescales in the active nucleus as well as
the timescales for ionization equilibration.
For typical AGN the transmission spectrum may not be well described by
models in ionization equilibrium, and the response to continuum variations
is smoothed over recombination timescales that could be, depending
on circumstances, anywhere from days to years.

\acknowledgments

    We thank Ski Antonucci, Chris Done, and Andy Fabian for stimulating
conversations and acquainting us with unpublished data.

    This work was partially supported by NASA Grants NAGW-3516 and NAG5-1630
and NASA contract NAS-5-27000 to the Johns Hopkins University.

\clearpage

\begin{table*}

\begin{center}

\begin{tabular}{l c c c c}

Line & Model a & Model b & Model c & Model d \\
\tableline
Ly$\alpha$ & 2.60 & 55.8 & 0.263 & 5.32 \\
CIV $\lambda$1549 & $2.78 \times 10^{-4}$ & 0.0199 & $2.95 \times 10^{-7}$ &
$1.36 \times 10^{-5}$ \\
NV $\lambda$1240 & $4.42 \times 10^{-4}$ & 0.215 & $4.60 \times 10^{-7}$ &
$1.19 \times 10^{-4}$ \\
OVI $\lambda$1034 & 0.0209 & 34.2 & $2.10 \times 10^{-5}$ & 0.0150 \\

\end{tabular}

\end{center}

\tablenotetext{}{Note: These optical depths are based on a total column density
of $10^{23}$ cm$^{-2}$ and thermal velocity widths. }

\caption{Optical Depths of UV Absorption Lines} \label{Table 1}

\end{table*}

\clearpage

\begin{figure} \label{Figure 1}
\caption{$\log(\nu F_\nu)$ vs. $\log(\nu)$ for our generic
spectrum (solid line) and the MF spectrum (dotted line).}
\end{figure}

\begin{figure} \label{Figure 2}
\caption{The radiative equilibrium curve of $T$ vs. $\Xi$ for
our generic spectrum (solid line) and the MF spectrum (dotted line).  The
letters mark the four points at which we have computed warm reflection region
models.}
\end{figure}

\begin{figure} \label{Figure 3}
\caption{Soft X-ray spectra emitted by the warm plasma,
in units of $L_{44} N_{23} (\Delta \Omega/\pi)$ erg s$^{-1}$ eV$^{-1}$.
The solid curve shows the total of emission lines and continua, smoothed by a
Gaussian of constant fractional width $\Delta \nu/\nu = 0.05$.
The dotted curve shows the portion due to bremsstrahlung and recombination
continua alone, with no smoothing.
Note that the vertical scale is compressed by a factor of four relative to
the horizontal scale.
The temperature and ionization parameter are given in the lower left
corner of each panel, and the letters correspond to the four models
identified in Fig. 2.}
\end{figure}

\begin{figure} \label{Figure 4}
\caption{The intrinsic emission spectrum in the ultraviolet
in units of $L_{44} N_{23} (\Delta \Omega/\pi)$ erg s$^{-1}$ \AA $^{-1}$.}
\end{figure}

\begin{figure} \label{Figure 5}
\caption{Reflected fraction as a function of frequency.}
\end{figure}

\begin{figure} \label{Figure 6}
\caption{Reflected plus intrinsic soft X-ray spectrum
in units of $L_{44} N_{23} (\Delta \Omega/\pi)$ erg s$^{-1}$ eV$^{-1}$.}
\end{figure}

\begin{figure} \label{Figure 7}
\caption{Transmission fraction in the soft X-ray band
as a function of
frequency, including those photons scattered into the observer's line of
sight from other directions.  The covering fraction of the absorbing
material is taken to be 1/4.
The transmission fraction for pure electron scattering is shown as a
dashed line.  Note that the zero has been suppressed on the vertical scale
to show the spectral features more clearly.}
\end{figure}

\end{document}